# A Novel Non-Stationary Bias-Correction Framework for Hurricane Risk Assessment under Climate Change

**Reda Snaiki[1], Teng Wu[2,*]**

[1] *Department of Construction Engineering, École de Technologie Supérieure, Université du Québec, Montréal, Canada*

[2] *Department of Civil, Structural and Environmental Engineering, University at Buffalo, Buffalo, NY 14260, USA*

[*] *Corresponding author. Email:* tengwu@buffalo.edu

**Abstract:** Conventional hurricane track generation methods typically depend on biased outputs from Global Climate Models (GCMs), which undermines their accuracy in the context of climate change. We present a novel non-stationary bias correction framework, developed for and applied to the North Atlantic basin, that adaptively accounts for evolving biases in GCM outputs. The framework combines dimensionality reduction with data-driven surrogate modeling to learn the nonlinear evolution between historical and projected GCM climate states, allowing the bias correction to evolve with changing climate conditions. The learned relationship is subsequently applied to a reanalysis-based historical baseline to generate bias-corrected environmental fields under future climate scenarios. Using the empirical Weibull plotting approach, we calculate return periods for wind speed and rainfall across selected North American coastal cities. Our results reveal substantial differences in projected hurricane hazards with and without bias correction, highlighting the importance of accounting for evolving GCM biases in climate-risk assessment and resilience planning. This work highlights the necessity of adaptive techniques for accurately assessing future climate impacts, offering a critical advancement in hurricane risk modeling and resilience planning.

## 1. INTRODUCTION

Hurricane-induced hazards pose significant threats to coastal communities. Characterized by intense winds, torrential rainfall, and storm surge, these events inflict substantial economic and human costs (Gori et al., 2023). Climate change is expected to alter tropical-cyclone characteristics, with projected increases in storm intensity and rainfall rates under warming conditions, while changes in overall storm frequency remain more uncertain (Knutson et al., 2020). This intensification, coupled with coastal urbanization and population growth, will likely lead to a nonlinear increase in hurricane-related risks (Olsen, 2015). To mitigate these growing threats and ensure the resilience of coastal regions, accurate estimation of hurricane-induced risk under current and future climate scenarios is imperative. Such assessments can inform the development of effective adaptation strategies, protect vulnerable populations, and guide the design and retrofitting of infrastructure (Snaiki and Parida, 2023b; Naeini et al., 2026).

Global Climate Models (GCMs) are essential tools for understanding future climate scenarios. However, their coarse resolution limits their ability to accurately simulate hurricanes, especially in terms of intensity and track (Knutson et al., 2020; Fiedler et al., 2021). While newer GCMs offer higher resolution (Haarsma et al., 2016), they still struggle to capture hurricane processes due to insufficient spatial resolution and reliance on parameterizations (Davis, 2018). Additionally, their limited temporal scope hinders probabilistic analysis of extreme hurricane events (Haarsma et al., 2016). To overcome the limitations of GCMs in simulating hurricane characteristics and their limited temporal scope, researchers have explored alternative approaches. Among these are statistical and statistical-dynamical hurricane track models, which enable the synthetic downscaling of hurricane activity and facilitate risk estimation without relying on detailed reanalysis or climate model simulations (Jing and Lin, 2020; Bloemendaal et al., 2020; Emanuel, 2021). These approaches have gained significant traction in recent years (Emanuel et al., 2006; Vickery et al., 2009b; Lin et al., 2012; Snaiki and Wu, 2020b, c; Snaiki and Parida, 2023a; Naeini et al., 2026). A key component of these methods is the creation of a large database of synthetic storms, encompassing their entire life cycle from genesis to dissipation. By leveraging information from a broader geographic area, this approach ensures sufficient data for estimating annual probabilities of low-frequency, high-impact events in hurricane-prone regions. These techniques typically involve three modules: genesis, translation, and intensity. While integrating GCM environmental parameters into the statistical-dynamical models allows for the consideration of future climate scenarios, this approach can inherit the biases inherent in the GCMs themselves (Gori et al., 2022). GCM biases, which refer to the systematic differences between the simulated and observed climate variables, arise from various factors including model simplifications, limited resolution, and inaccurate representation of complex processes like cloud formation and ocean-atmosphere interactions. These biases can manifest in multiple ways, such as mean, variance, spatial, temporal, and extreme event biases. GCM biases are particularly problematic for climate change impact assessments, as they can distort the understanding of future climate conditions, especially for extreme events like hurricanes. To mitigate these issues, bias correction methods are applied to GCM outputs to reduce systematic errors and improve the alignment of simulations with observed historical data.

Bias-correction techniques for GCMs can be broadly categorized into four main approaches: Quantile Mapping, Delta Method, Statistical Downscaling, and Machine Learning. Quantile Mapping aligns the cumulative distribution function (CDF) of model output with the CDF of observations, preserving the overall distribution shape while correcting biases in mean and variance (Cannon et al., 2015). The Delta Method is a simpler approach that adjusts model output by constant or time-varying factor (Navarro-Racines et al., 2020; Bloemendaal et al., 2022). While easy to implement, it may not capture complex climate system changes. Statistical Downscaling links large-scale GCM output to local-scale observations (Ahmed et al., 2013; Tabari et al., 2021). This technique can generate high-resolution climate projections, including those relevant to hurricane genesis and intensity. Machine Learning models, like artificial neural networks and support vector machines, can be trained on historical observations and model output to learn complex relationships and correct biases (Zhang et al., 2024). While various bias-correction techniques have been explored (Murakami et al., 2014; Gori et al., 2022), most focus on mapping GCM outputs to observational data to reduce current model biases. These models are typically trained on historical data and applied to future GCM outputs, assuming that biases remain consistent over time. However, this assumption may be limiting, as biases themselves can evolve over time. Conversely, some studies (Tabari et al., 2021; Bloemendaal et al., 2022) have attempted to dynamically correct GCM biases through multiplicative or additive adjustments. However, these approaches are fundamentally linear and may struggle to capture the highly nonlinear dynamics inherent in GCM outputs. Therefore, advanced techniques should be explored that can explicitly learn how biases might change under different climate scenarios.

This study proposes a novel data-driven bias-correction framework for climate-model environmental fields under changing climate conditions. The approach is designed to account for the temporal evolution of model bias by learning the nonlinear relationship between historical and projected GCM climate states. Given the high dimensionality of GCM outputs, dimensionality reduction is first applied to the historical and projected fields from the NOAA Geophysical Fluid Dynamics Laboratory CM4.0 model [GFDL-CM4 (Held et al., 2019)] to extract the dominant spatial patterns and their corresponding temporal coefficients. A reduced-order surrogate model is then trained to relate the historical GCM state to its projected counterpart. Once identified, this relationship is applied to a reanalysis-based historical state to generate bias-corrected environmental fields under projected climate conditions. The transferability of the learned relationship is assessed through a historical hindcast for which the corresponding reanalysis data are independently available. Finally, the corrected environmental fields are incorporated into a physics-based hurricane simulation framework to quantify their influence on wind-speed and rain-rate return periods at selected North American coastal locations.

## 2. METHODS

### 2.1 Hurricane track methodology

The downscaling method leverages information from a broader geographic region to populate the target area with statistically robust data, enabling the estimation of the annual probability of low-frequency, high-impact hurricane events. The core components of this track model typically include three key modules: genesis, translation, and intensity. In this study, a physics-informed hurricane track model was employed to generate synthetic tracks (Emanuel et al., 2008; Emanuel, 2017; Lin et al., 2023). This model outperforms traditional hurricane track models, which often rely on simplified regression formulas and may struggle to capture the complex, non-linear relationships within hurricane data. The genesis model employed in this study adopts a stochastic approach, randomly seeding potential storm locations across both space and time (Emanuel et al., 2008; Lin et al., 2020; Emanuel, 2022). These seeded disturbances are subsequently allowed to evolve and interact with the ambient large-scale environment. To ensure a realistic spatial distribution and prevent genesis near the equator where the Coriolis force is insufficient, the random seeding is weighted by a function that depends on the background vorticity and latitude, similar to the approach used in genesis potential indices (Lin et al., 2023). Following the methodology of Lin et al. (2023), each seed is initialized with a weak intensity (5 m/s). The model then tracks its evolution for two days. Only those seeds that survive and intensify to a minimum threshold, defined as a maximum wind speed of 18 m/s, are retained and considered to have undergone genesis. This process effectively simulates a wide range of observed hurricane formation patterns.

Following the introduction of storm seeds, the storm trajectory is modeled using the beta-and-advection framework (Emanuel et al., 2006; Lin et al., 2020). This model posits that the storm's trajectory is primarily driven by the interaction of large-scale wind fields and a systematic poleward and westward drift (Emanuel et al., 2006). The translational velocity ($\mathbf{v}_t$) of the storm is calculated using the following equation (Emanuel et al., 2006; Lin et al., 2020):

$$\mathbf{v}_t = (1-\alpha)\mathbf{v}_{250} + \alpha\mathbf{v}_{850} + \mathbf{v}_\beta \cos(\phi) \quad (1)$$

where $\mathbf{v}_{850}$ = large-scale environmental wind at 850-hPa; $\mathbf{v}_{250}$ = large-scale environmental wind at 250-hPa; $\mathbf{v}_{\beta}$ = translational speed correction; $\phi$ = latitude; and $\alpha$ = steering coefficient. A key feature of the model employed in this study is the implementation of an intensity-dependent steering flow, following the findings of Lin et al. (2023). The steering coefficient, $\alpha$, is not treated as a constant; instead, it varies linearly with the storm's intensity. This allows the model to simulate the important physical process of stronger, deeper storms being steered by higher-altitude winds, which provides a more realistic representation of the storm's trajectory.

The intensity evolution along the storm's track is simulated using the FAST intensity model, a well-established framework based on a simplified pair of coupled, non-linear ordinary differential equations (Emanuel, 2017; Emanuel and Zhang, 2017). The model evolves two key variables: the maximum azimuthal wind speed ($v$) and a non-dimensional inner-core moisture variable ($m$), which can be conceptualized as a type of relative humidity. The model is forced by the large-scale environment, and its governing equations are expressed as (Emanuel, 2017; Emanuel and Zhang, 2017):

$$\frac{dv}{dt} = \frac{1}{2}\frac{C_k}{h}\left[\alpha_o \beta V_p^2 m^3 - (1-\gamma m^3)v^2\right] \tag{2a}$$

$$\frac{dm}{dt} = \frac{1}{2}\frac{C_k}{h}\left[(1-m)v - \chi S m\right] \tag{2b}$$

$$\beta = 1 - \epsilon - \kappa \tag{2c}$$

$$\gamma = \epsilon + \alpha_o \kappa \tag{2d}$$

$$\epsilon = \frac{T_s - T_0}{T_s} \tag{2e}$$

$$\kappa = \frac{\epsilon}{2}\frac{C_k}{C_d}\frac{L_v}{R_d}\frac{q_s^*}{T_s} \tag{2f}$$

$$\alpha_o = 1 - 0.87 exp^{-z} \tag{2g}$$

$$z = 0.01\Gamma^{-0.4} h_m u_T V_p v^{-1} \tag{2h}$$

$$\chi = \frac{s^* - s_m}{s_0^* - s^*} \tag{2i}$$

where $C_k$ = surface enthalpy exchange coefficient; $C_d$ = drag coefficient; $h$ = boundary layer height; $V_p$ = potential intensity; $\alpha_o$ = ocean interaction parameter; $S$ = the 250-850-hPa vertical wind shear; $\chi$ = gridded mid-level saturation entropy deficit; $s^*$ = saturation moist entropy of the free troposphere; $s_0$ = saturation moist entropy of the sea surface; $s_m$ = moist entropy of the middle troposphere; $T_s$ = surface temperature; $T_0$ = outflow temperature; $L_v$ = constant latent heat of vaporization; $R_d$ = dry gas constant; $q_s^*$ = surface saturation specific humidity; $\epsilon$ = thermodynamic efficiency; $\Gamma$ = sub-mixed layer thermal stratification; $h_m$ = mixed layer depth; and $u_T$ = translation speed. As shown by the governing equations, the key environmental quantities that drive the intensification process are the potential intensity $V_p$, vertical wind shear $S$, and the environmental entropy deficit $\chi$, along with various ocean and surface thermodynamic properties.

Following hurricane track generation, wind and rain hazards are simulated. Wind speeds are determined using an analytical model (Snaiki and Wu, 2017), while rainfall intensities are estimated based on an empirical model (Tuleya et al., 2007). These hazard simulations are then coupled with the generated hurricane tracks. Since hurricane-track modules require inputs of environmental predictors [e.g., sea surface temperature (SST)], the GCM fields (of environmental predictors) are first bias-corrected and then used to drive the tracks within the North Atlantic basin.

**2.2 Data-driven bias-corrected framework**

Several bias-correction techniques, such as quantile mapping, delta method, statistical downscaling, and machine learning, are commonly used to address biases in GCMs and improve their applicability for localized studies like hurricane intensity and precipitation projections. However, most existing techniques focus on mapping current or historical GCM outputs to historical observations, assuming that future biases will remain similar to past ones. In addition, the existing methods for dynamically correcting GCM biases often rely on linear adjustments, such as multiplicative or additive factors. These approaches may have limitations in accurately capturing the non-linear characteristics inherent in GCM outputs. This study introduces a machine-learning framework to account for the non-stationarity of climate-model biases. The approach learns the nonlinear relationship between historical and projected GCM climate states in a reduced-order space and subsequently transfers this relationship to a reanalysis-based historical state. In this manner, the projected environmental fields incorporate the climate-change signal represented by the GCM while being conditioned on a historical reanalysis baseline. This formulation enables the bias correction to evolve with the projected climate state and avoids prescribing a fixed historical correction throughout the projection period. Algorithm 1 summarizes the proposed framework.

**Algorithm 1.** Algorithm: Data-Driven Bias-Corrected Framework

**Input**:

- Historical GCM data; projected GCM data; historical ERA5 reanalysis data.

**Step 1:** Dimensionality reduction using POD

1. Compute singular value decomposition (SVD):
   Perform SVD on the historical and projected data from the GCM model to extract the POD modes.

$$GCM_{Historical} = U_1 S_1 V_1^T$$
$$GCM_{Future} = U_2 S_2 V_2^T$$

2. POD mode selection:
   Select the first $K$ POD modes from $U_1$ and $U_2$ that capture at least 95% of the total variance in the data.

$$K = min\left\{k: \sum_{i=1}^{k} S_i^2 \Big/ \sum_{i=1}^{N} S_i^2 \geq 95\%\right\}$$

3. Projection onto reduced subspace:
   Project both the historical and future data onto the reduced subspace spanned by the selected $K$ POD modes to obtain the corresponding time-dependent POD coefficients.
   - For historical data:

$$a^H = U_{1,K}^T \times GCM_{Historical}$$

   - For projected data:

$$a^P = U_{2,K}^T \times GCM_{Future}$$

**Step 2:** Reduced-order climate-state mapping

4. Train reduced-order surrogate:
   Use the historical POD coefficients $a^H$ as input and the corresponding future POD coefficients $a^P$ as output to train the reduced-order surrogate. The first POD coefficient is modeled using linear regression, while the remaining retained coefficients are modeled using an LSTM network.

$$f_\theta: a^H \rightarrow a^P$$

5. Projection of ERA-5 data:
   Project the historical ERA-5 data onto the POD basis obtained from $GCM_{Historical}$ to obtain the corresponding reduced coefficients:

$$a^{ERA5} = U_{1,K}^T \times ERA5$$

6. Prediction of future POD coefficients:
   Apply the trained reduced-order surrogate to the ERA5 coefficients to obtain the corresponding projected POD coefficients:

$$a^{future} = f_\theta(a^{ERA5})$$

**Step 3:** Reconstruction of future data field

7. Reconstruction of future data:
   Reconstruct the projected data field using the predicted POD coefficients $a^{future}$ and the POD modes obtained from the future GCM scenario.

$$ERA5_{future} = U_{2,K} \times a^{future}$$

**Output:**

- The reconstructed environmental field under future climate conditions, $ERA5_{future}$, conditioned on the historical ERA5 state.

---

As outlined in Algorithm 1, the framework comprises two core modeling components, dimensionality reduction and reduced-order climate-state mapping, followed by reconstruction of the projected field:

1) Dimensionality reduction. Given the high dimensionality of GCM output data, which presents significant challenges for both computational efficiency and model interpretability, a crucial initial step involves dimensionality reduction. Techniques such as wavelet-domain projection, dynamic mode decomposition, and Proper Orthogonal Decomposition (POD), can be used for this purpose. In this study, POD is employed to reduce the system's dimensionality by identifying the dominant spatial patterns in the data. POD decomposes the spatio-temporal data field into a set of orthogonal spatial basis vectors, known as POD modes, and a corresponding set of time-dependent coefficients. These modes are optimal in the sense that they capture the maximum possible variance (or energy) of the system for any given number of modes. Practically, these modes are computed using the Singular Value Decomposition (SVD) of the data matrix, as outlined in Algorithm 1. The columns of the left singular vectors ($U$) from the SVD form the spatial POD modes. The original high-dimensional data can then be represented by projecting it onto a small, truncated set of these modes ($K$ modes), resulting in $K$ time-series of POD coefficients. This process significantly reduces the complexity of the data fed to the subsequent machine learning model while retaining over 95% of the total variance of the original field, ensuring that the most important features of the climate dynamics are preserved. Separate POD decompositions are performed for the historical and projected GCM fields. The historical ERA5 field is subsequently projected onto the POD basis obtained from the historical GCM field, as described in Algorithm 1.

2) Reduced-order climate-state mapping. The relationship between the historical and projected reduced states is represented using a decoupled surrogate model. The dominant first POD coefficient is modeled using linear regression, while the remaining higher-order coefficients are modeled using an LSTM network. Collectively, the surrogate defines the mapping $f_\theta: a^H \rightarrow a^P$ between the historical and projected GCM states in the reduced-order space.

Once the mapping function $f_\theta$ is identified, it is applied to the historical ERA5 data in the reduced-order space. Specifically, the ERA5 field is projected onto the POD basis obtained from the historical GCM scenario to determine the corresponding coefficients $a^{ERA5}$. These coefficients are then supplied to the trained surrogate to obtain the projected coefficients $a^{future} = f_\theta(a^{ERA5})$. The projected environmental field is subsequently reconstructed using $a^{future}$ and the POD basis obtained from the future GCM scenario. In this way, the historical state is constrained by reanalysis data, while its evolution toward the projected state follows the relationship learned from the GCM. The framework is applied to the environmental parameters required for hurricane simulation, including monthly averaged sea surface temperature, mean sea level pressure, temperature and specific humidity at various pressure levels, and potential intensity. A schematic illustration of the approach is provided in Fig. 1.

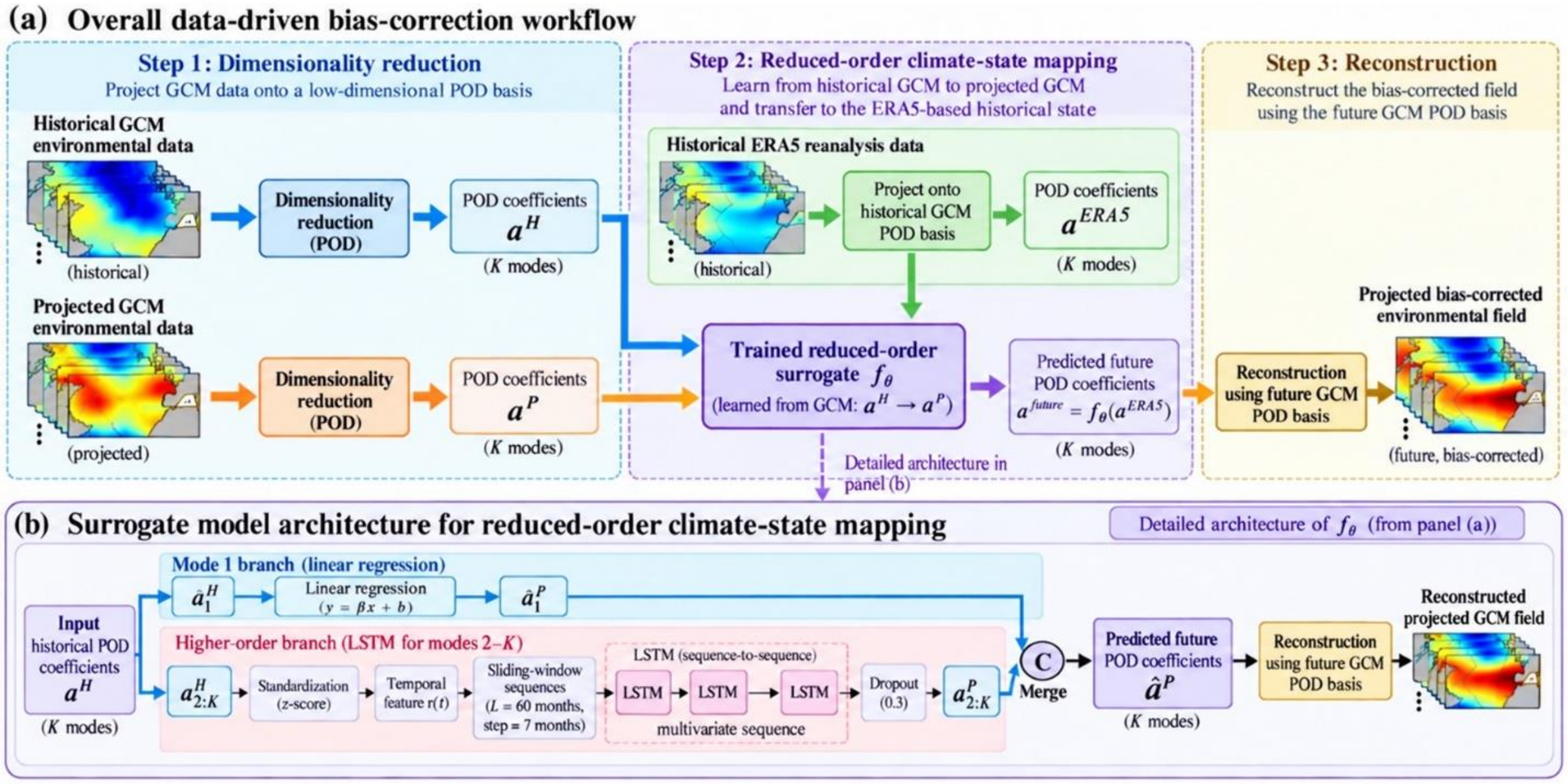

**Fig. 1.** Schematic illustration of the proposed bias-correction framework. (a) Overall procedure for learning the relationship between historical and projected GCM states and applying it to the historical reanalysis state. (b) Architecture of the reduced-order surrogate, in which the first POD coefficient is modeled using linear regression and the higher-order coefficients are modeled using an LSTM network.

It should be noted that while our framework is designed to account for biases that evolve over time, it is a statistical mapping learned offline. This is distinct from traditional dynamic bias correction methods where adjustments are applied in real-time during a model's operation. The novelty of the approach therefore lies in using a data-driven model to learn the nonlinear relationship between different GCM climate states, enabling the bias correction to evolve with the projected climate conditions.

## 2.3 Simulation scenarios

The bias-corrected environmental fields generated by the framework are used to assess hurricane risk and its evolution at six selected locations: Galveston (-94.79°; 29.29°), New Orleans (-90.04°; 29.82°), Miami (-80.12°; 25.79°), Myrtle Beach (-78.87°; 33.69°), Atlantic City (-74.49°; 39.38°) and Halifax (-63.59°; 44.65°). While the bias-correction model was trained on the full spatial domain of the North Atlantic basin, the corrected environmental fields are used to assess hurricane risk and its evolution at these six selected locations to demonstrate the bias-correction impact on local risk assessment. Five synthetic hurricane datasets, each spanning 10,000 years, were generated to assess hurricane hazard under the historical and projected climate conditions. These consist of one historical ERA5-based dataset and, for each future period, datasets generated using both the original GFDL-CM4 fields and the corresponding bias-corrected fields. The datasets therefore represent three climate periods:

- Historical Baseline (1979-2014). Based on the ERA5 reanalysis data, this scenario provides a reference point for comparison.
- Near-Future Scenario (2024-2059). Two synthetic catalogs are generated under SSP5-8.5: one using the original GFDL-CM4 environmental fields and one using the corresponding bias-corrected fields produced by the proposed framework.
- Far-Future Scenario (2060-2095). Two analogous synthetic catalogs are generated using the original and bias-corrected GFDL-CM4 environmental fields under SSP5-8.5.

The 10,000-year duration refers to the stochastic hurricane catalog generated under each prescribed climate state, rather than to the duration of the underlying climate projection. Long synthetic catalogs are well established in probabilistic tropical-cyclone hazard assessment and provide the sample size required to estimate low-probability hazards that cannot be resolved directly from relatively short observational or climate-model records (Emanuel et al., 2006; Vickery et al., 2009b; Lin et al., 2023).

The choice of 1979 as the starting year for the historical baseline aligns with the beginning of the satellite era, providing a more reliable and consistent observational record. The two future climate periods, 2024–2059 and 2060–2095, allow the assessment of both near- and longer-term changes in hurricane hazard. The resulting synthetic hurricane tracks provide hourly time series of storm characteristics, including longitude, latitude, and maximum wind speed. These tracks are subsequently coupled with wind and rainfall hazard models (Snaiki and Wu, 2017; Tuleya et al., 2007) to estimate the corresponding intensity measures and return periods. Although these models are adopted in the present study, other established wind-field models (Kepert, 2001; Vickery et al., 2009a; Snaiki and Wu, 2020a) and rainfall models (Langousis and Veneziano, 2009; Snaiki and Wu, 2019; Xi et al., 2020) could also be integrated within the framework.

### 2.4 Return periods

In this study, the return period, also known as the mean recurrence interval (MRI), has been used to analyze the wind speed and rain rate in the context of climate change. This is a common approach but has known limitations, especially when dealing with nonstationary data. However, there is no consensus on how to best handle nonstationarity in return period calculations, especially for wind speeds under climate change. Many studies still rely on return period due to its simplicity and established interpretability. Furthermore, since the focus is on comparing different climate change scenarios rather than absolute predictions, return period provides a consistent basis for relative comparisons, even if it does not fully capture nonstationarity. In addition, two periods for the simulation of future climate scenarios (i.e., 2024-2059 and 2060-2095) were selected to further limit the effects of nonstationarity.

Two primary approaches are commonly used to estimate return periods: (i) empirical plotting-position methods, which avoid assuming a specific probability distribution; and (ii) parametric extreme-value methods, in which a selected probability distribution is fitted to the data. This study adopts the empirical Weibull plotting-position approach (Weibull, 1939) to estimate return periods for the selected hurricane hazard intensity measures, namely maximum wind speed and rain rate. The estimation is conducted on the five synthetic hurricane datasets each comprising 10,000 years of synthetic hurricanes. The long synthetic record is required by the empirical nature of the adopted plotting-position approach. Because the return levels are estimated directly from the simulated event population rather than extrapolated from a fitted extreme-value distribution, a short catalog would provide insufficient sampling of rare events. In particular, a 100-year synthetic record would not adequately support the estimation of the 500-, 1000-, and longer return periods considered in this study. The Weibull plotting-position approach facilitates a straightforward calculation of the return period. For instance, the return period for wind speed can be calculated using the following expression (Makkonen, 2006):

$$RP(v) = \frac{1}{P_e(v)} = \frac{n+1}{i} \cdot \frac{m}{n} \tag{3}$$

where $P_e(v)$ = exceedance probability for a given maximum wind speed $v$ at rank $i$; $n$ = number of storms events in the synthetic database; and $m$ = length in years of the dataset (here $m$ = 10,000).

# 3. RESULTS

## 3.1. Model validation

In this study, the framework is trained, validated, and applied exclusively to the North Atlantic basin. This section details the performance and validation of the proposed non-stationary bias correction framework. The framework was applied to all environmental variables required by the hurricane track model, with comparable successful results obtained for each. For clarity and concision, SST is used as the representative parameter for the detailed validation since it is a fundamental driver of hurricane intensity and its high spatio-temporal variance provides a robust and challenging test case for the framework's ability to learn and map complex, evolving climate patterns. The validation proceeds in two stages: (i) training-domain performance within the GCM’s simulated environment for near-future and far-future scenarios, and (ii) a historical hindcast against ERA5 to establish applicability to real-world conditions.

### 3.1.1. Model training and performance on GCM data

The initial phase of the methodology focuses on training the machine learning (ML) framework and evaluating its performance entirely within the GCM's simulated environment. To account for different projection horizons, the framework is trained independently for two separate mappings: one from the historical GCM period (1979-2014) to a near-future GCM period (2024-2059), and a second to a far-future GCM period (2060-2095).

The training process begins with extensive data pre-processing. For a given environmental parameter, the raw GCM data often contains masked or invalid grid points (e.g., land points for SST) encoded as extreme constants. These values are systematically replaced with NaN to avoid numerical contamination. The 3D GCM data cube (with dimensions time × latitude × longitude) is then reshaped into a 2D matrix (spatial points × time). A master mask is created from the data to identify and filter out these invalid points, resulting in a final matrix where each row represents the complete time series for a single valid grid point. POD is applied separately to both historical and future matrices. This yields a set of spatial POD modes and corresponding temporal coefficients. These temporal coefficients represent the dominant time-varying structures in the data.

We adopt a decoupled modeling strategy to capture the dynamics of the different POD modes. The first POD coefficient is modeled using linear regression. The remaining retained higher-order coefficients are modeled using an LSTM neural network. The LSTM architecture consists of one hidden layer with 69 units and a dropout rate of 0.3 and is trained using the Adamax optimizer with a learning rate of 0.001 and a batch size of 37. The training procedure for the LSTM is as follows. First, the higher-order historical and projected GCM POD coefficients are standardized prior to sequence construction. For $j = 2, \dots, K$, $\tilde{a}_j^H(t) = \frac{a_j^H(t) - \mu_j^H}{\sigma_j^H}$, $\tilde{a}_j^P(t) = \frac{a_j^P(t) - \mu_j^P}{\sigma_j^P}$ where $\mu_j^H$, $\sigma_j^H$, $\mu_j^P$, and $\sigma_j^P$ denote the corresponding means and standard deviations used to standardize the historical and projected coefficient series. Denoting the standardized higher-order historical coefficients at time $t$ by $\tilde{\boldsymbol{a}}^H(t) = [\tilde{a}_2^H(t), \dots, \tilde{a}_K^H(t)]^T$, the LSTM input feature vector is written as $\boldsymbol{z}(t) = [\tau(t), \tilde{\boldsymbol{a}}^H(t)^T]^T$, where $\tau(t)$ denotes the scalar temporal feature appended to each time step. The $i$-th input sequence is then $\mathbf{X}_i = [\boldsymbol{z}(t_i), \boldsymbol{z}(t_i + 1), \dots, \boldsymbol{z}(t_i + L - 1)]^T$, while the corresponding target sequence is $\mathbf{Y}_i = [\tilde{\boldsymbol{a}}^P(t_i), \tilde{\boldsymbol{a}}^P(t_i + 1), \dots, \tilde{\boldsymbol{a}}^P(t_i + L - 1)]^T$, where $\tilde{\boldsymbol{a}}^P(t) = [\tilde{a}_2^P(t), \dots, \tilde{a}_K^P(t)]^T$. Accordingly, $\mathbf{X}_i \in \mathbb{R}^{L \times K}$, $\mathbf{Y}_i \in \mathbb{R}^{L \times (K-1)}$. Here, $L = 60$ months is the sequence length, and successive sequences are initiated with an increment of $s = 7$ months. This sequence-based setup allows the LSTM to learn temporal dependencies and nonlinear relationships across the POD modes. The trained LSTM model outputs predicted future POD coefficients, which are then inverse-transformed to recover their original scale. Finally, the predicted coefficients from the linear model (Mode 1) and the LSTM (Modes 2–K) are combined and mapped back to the original physical space using the POD basis vectors from the future GCM scenario. The result is the final bias-corrected field for the environmental parameter, reconstructed on the full spatial grid with invalid points restored as NaN to maintain consistency with the original GCM data format.

The GCM-domain performance was strong for both scenarios. Figures 2 and 3 display spatial snapshots of the target SST, the model's prediction, and the absolute error for selected years in the testing sets of both the near-future (2052, 2054, 2056) and far-future (2088, 2090, 2092) scenarios, respectively. The high degree of visual similarity and low error magnitudes confirm the model’s ability to reconstruct the complex spatial fields accurately.

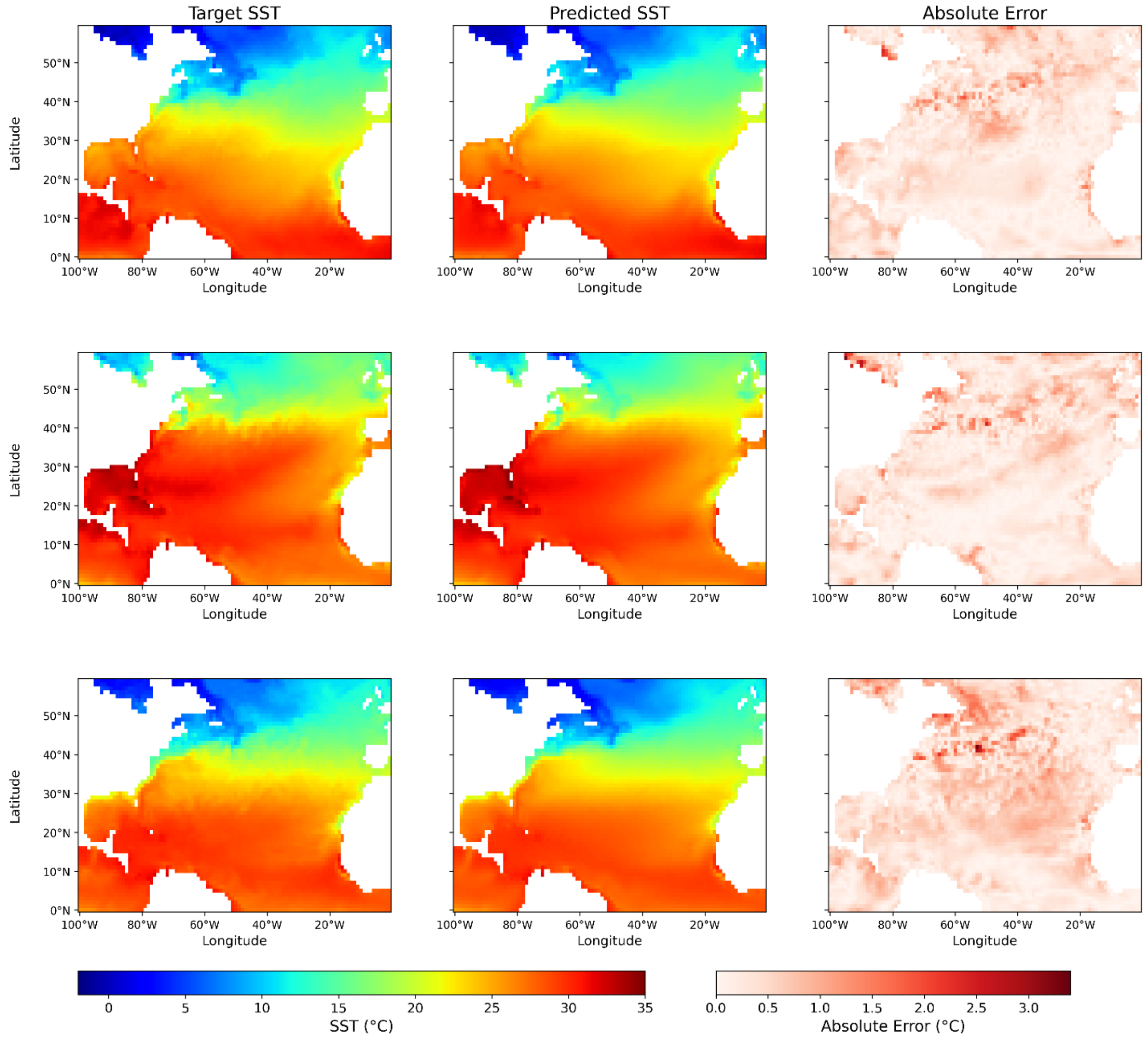

**Fig. 2.** GCM training performance snapshots for the near-future scenario over the North Atlantic basin. Each column shows, from left to right, the target SST, predicted SST, and the absolute error between them. Each row corresponds to a different time within the testing period: 05/2052 (top), 08/2054 (middle), and 11/2056 (bottom).

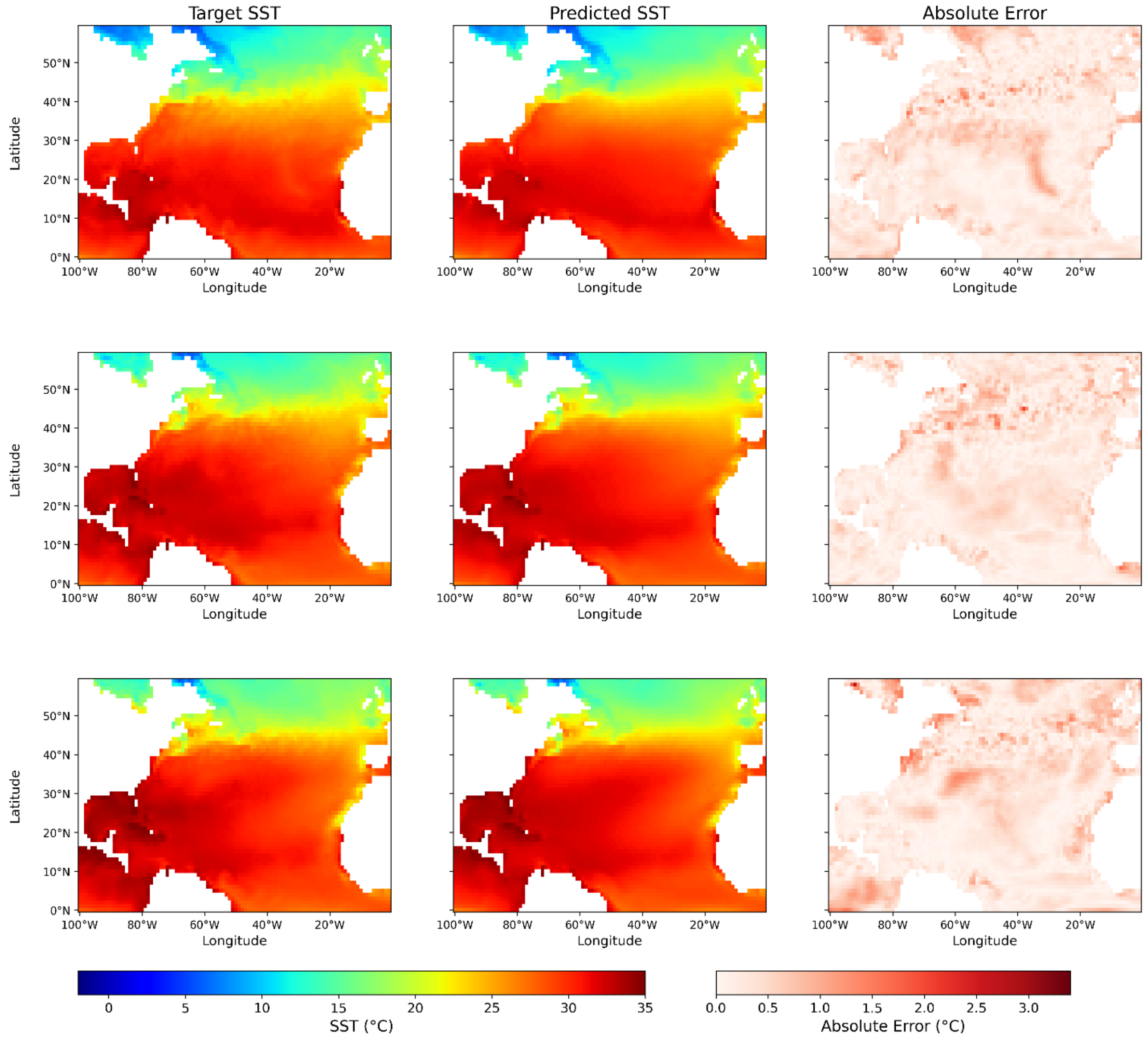

**Fig. 3.** GCM training performance snapshots for the far-future scenario over the North Atlantic basin. Each column shows, from left to right, the target SST, predicted SST, and absolute error between them. Each row corresponds to a different time within the testing period: 10/2088 (top), 09/2090 (middle), and 08/2092 (bottom).

To provide a more robust assessment of the spatial performance beyond a few snapshots, the time-averaged absolute error was computed for each grid point over the entire simulation period. Figure 4 presents these spatial error maps for both the near-future and far-future scenarios. The results show that the mean absolute error remains low over most of the North Atlantic basin, with larger errors confined to relatively localized regions and maximum values not exceeding approximately 1.4 °C. This indicates that the framework maintains good spatial fidelity over both the near-future and far-future testing periods.

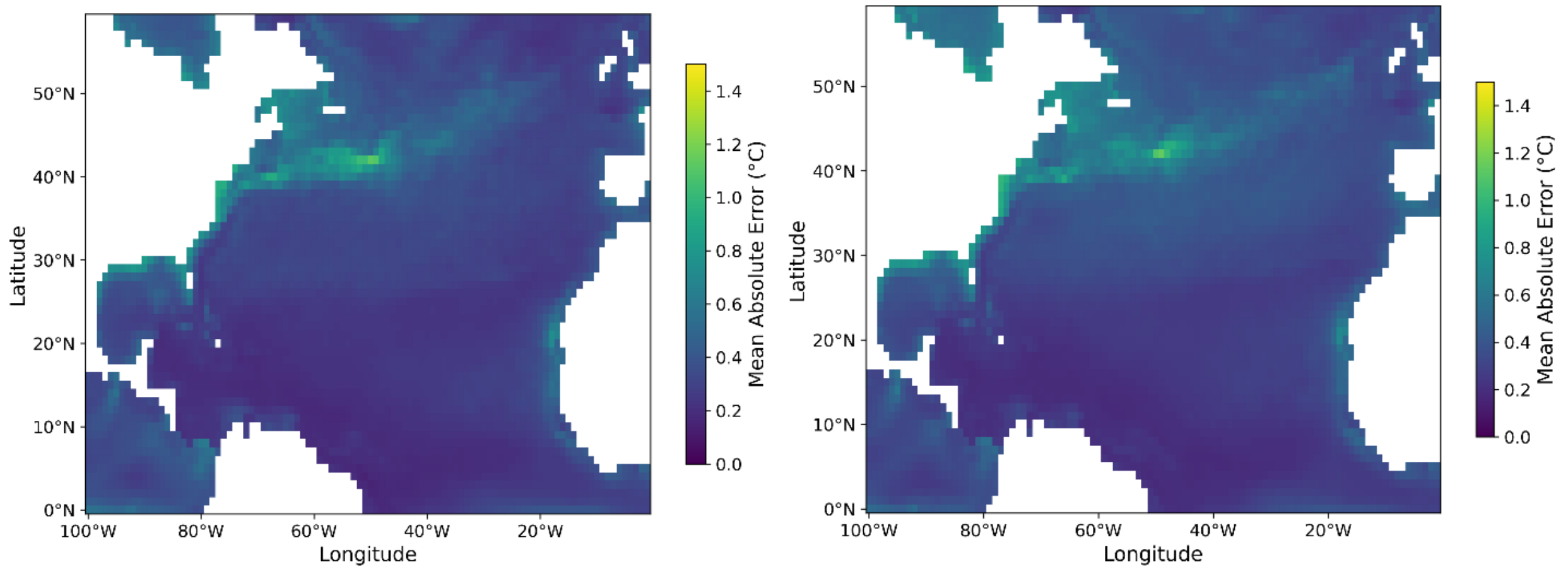

**Fig. 4.** Time-averaged absolute spatial error for the GCM training performance for the (a) near-future (2024-2059) and (b) far-future (2060-2095) scenarios.

This qualitative assessment is supported by quantitative metrics, as shown in Fig. 5, which shows the evolution of the RMSE over both the near-future and far-future periods. The plots clearly distinguish between the training RMSE (blue) and the testing RMSE (orange). In both scenarios, the testing error remains low and comparable in magnitude to the training error, indicating stable predictive performance over the evaluation period. The overall training (testing) RMSE for the near-future mapping was 0.41 (0.51) °C, while the far-future mapping yielded an RMSE of 0.44 (0.53) °C.

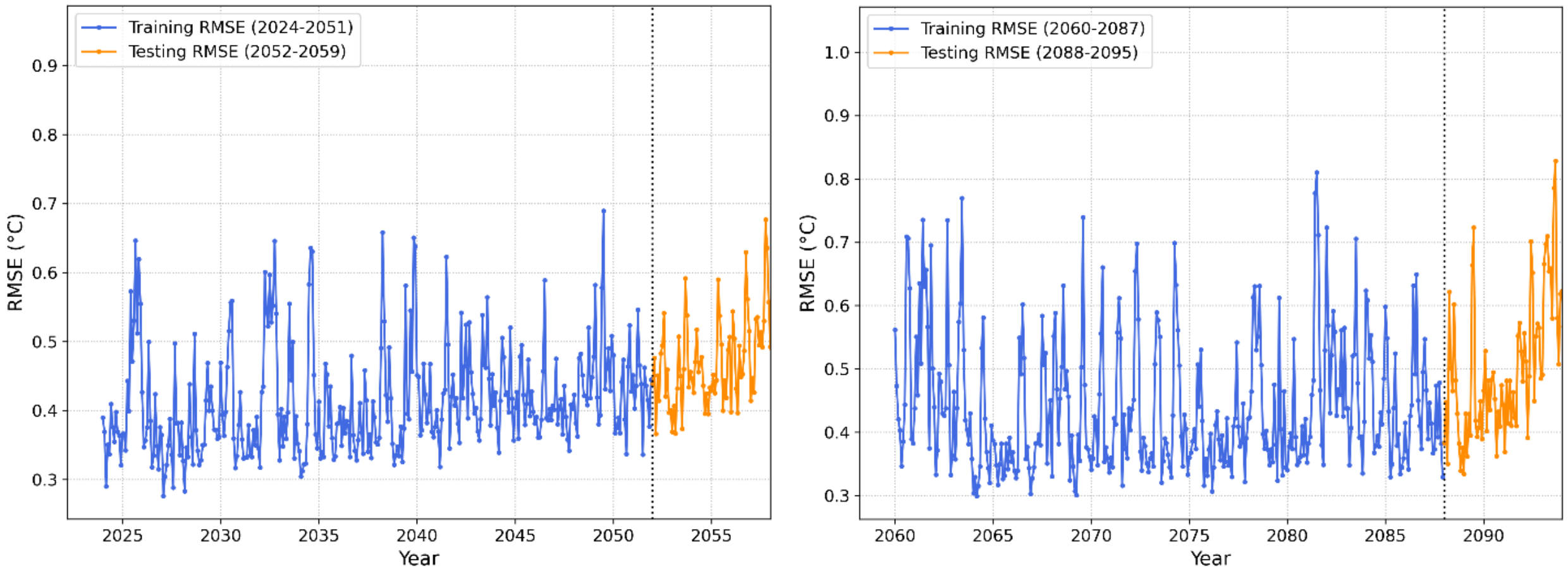

**Fig. 5.** Temporal RMSE for the training and testing subsets of the near-future (left) and far-future (right) mappings.

To further quantify the model's performance and directly assess the multi-year mean bias and error distribution, a histogram of the prediction errors (Predicted - Target) was generated. Figure 6 shows these error probability distribution functions (PDFs) for both the near-future and far-future scenarios. The distributions are narrow and centered almost perfectly around zero, with the mean bias being negligible in both cases. This indicates that the models exhibit negligible overall mean bias and approximately symmetric error distributions for both future scenarios, providing additional evidence of stable predictive performance. Furthermore, the performance in the reduced-order space was examined. As shown in Fig. 7, the predicted temporal coefficients for representative modes (e.g., Modes 3 and 4) closely track the target coefficients for both future scenarios. This demonstrates that the LSTM successfully captured the non-linear temporal dynamics of the system's key fluctuation patterns.

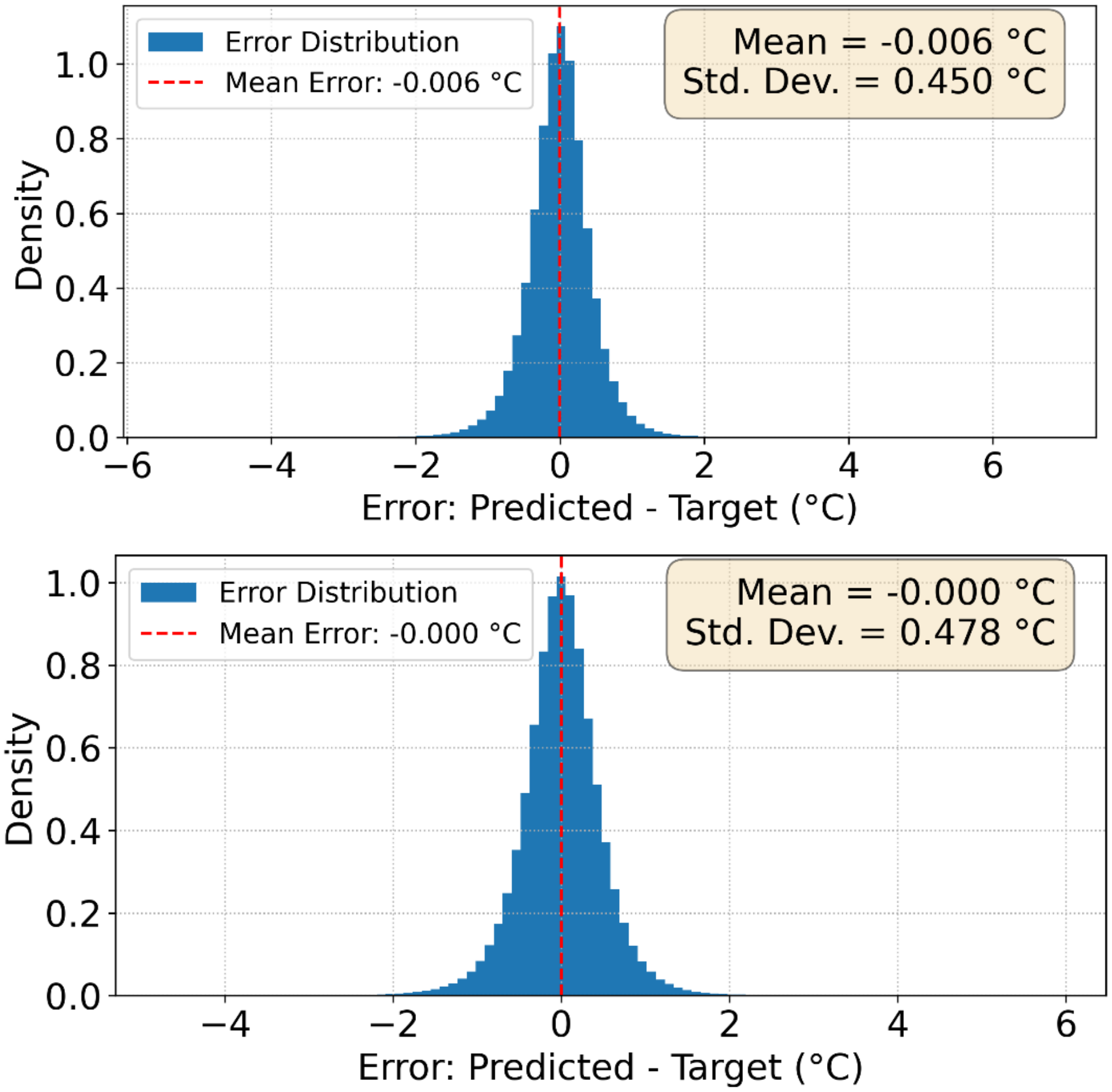

**Fig. 6.** Probability distribution functions of the prediction error (Predicted - Target) from the GCM training for the (a) near-future (2024-2059; upper plot) and (b) far-future (2060-2095; bottom) scenarios.

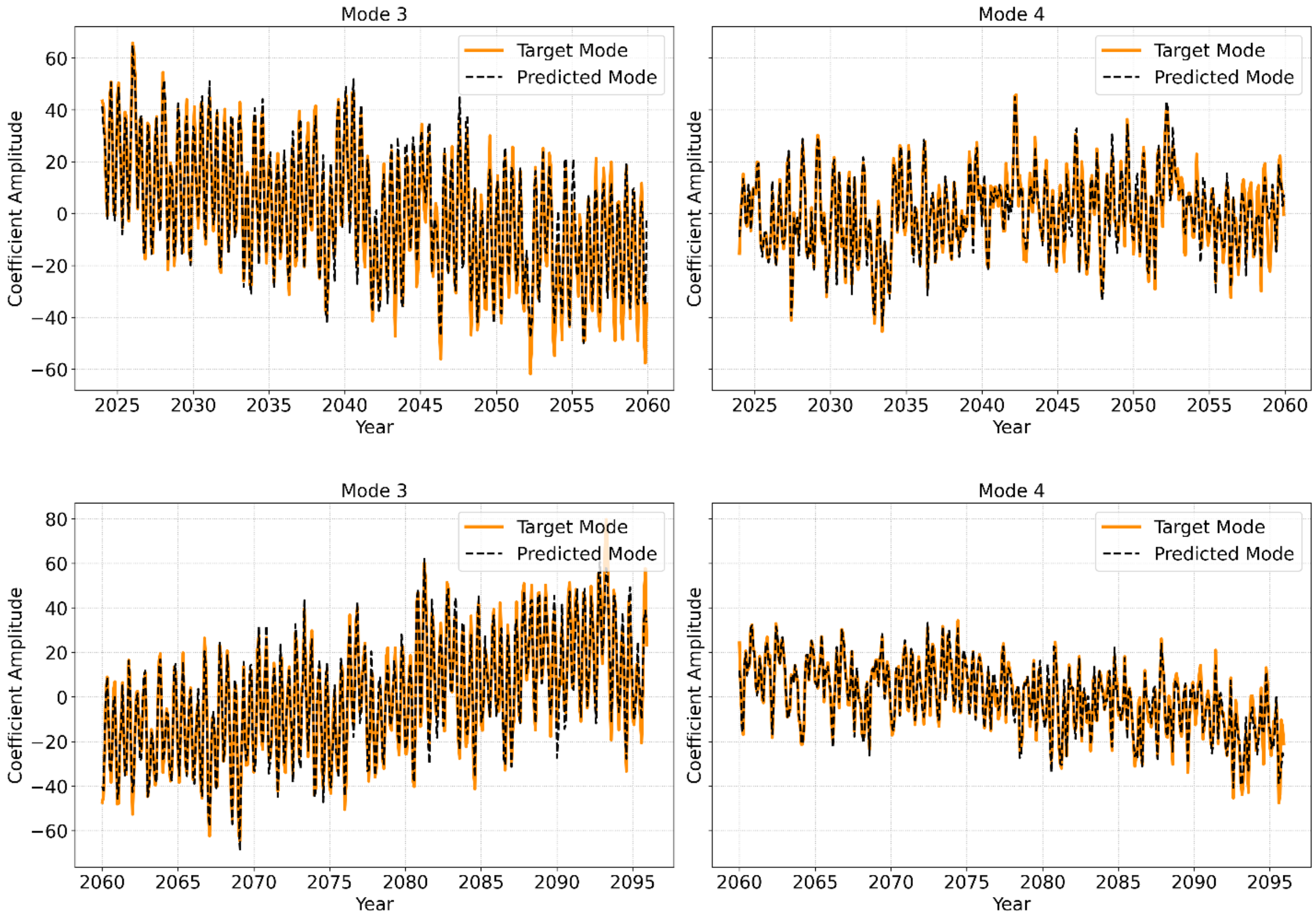

**Fig. 7.** POD temporal coefficient comparison for the near-future (top row) and far-future (bottom row) scenarios.

It is noted that the framework's performance was validated for all key environmental parameters with comparable success and the results are briefly summarized here. Specifically, the analysis of the prediction error probability density functions for other variables revealed similarly negligible biases and narrow distributions. For the sea-level pressure, the model achieved a mean error of -0.009 hPa (standard deviation, $\sigma$ = 2.147 hPa) for the near-future and -0.006 hPa ($\sigma$ = 2.028 hPa) for the far-future scenario. For specific humidity, the mean errors were low at $-2.03 \times 10^{-7} kg/kg$ ($\sigma$ =$3.125 \times 10^{-4} kg/kg$) for the near-future and $1.007 \times 10^{-6} kg/kg$ ($\sigma$ =$3.692 \times 10^{-4} kg/kg$) for the far future. For air temperature, the framework yielded mean errors of -0.024 °C ($\sigma$ = 0.949 °C) and -0.001 °C ($\sigma$ = 1.023 °C) for the near and far-future periods, respectively. Comparable near-zero biases and stable dispersion are observed for the remaining environmental drivers (e.g., winds at 850/250 hPa), supporting the robustness of the correction step prior to the hurricane-risk analysis.

#### 3.1.2. Historical validation against ERA5 reanalysis

A key test of the framework is its transfer to reanalysis data through a historical hindcast experiment. For this validation, a model was trained exclusively on the GCM's simulated evolution from its 1979-1996 period to its 1997-2014 period. This trained model was then tasked with predicting the ERA5 climate state for 1997-2014 by using only the ERA5 reanalysis data from 1979-1996 as input. The resulting prediction was then compared against the ERA5 reference field for 1997-2014. This hindcast provides an independent test of whether the relationship learned from the GCM remains informative when applied to the reanalysis-based historical state.

The results provide evidence of the framework's historical transferability. Figure 8 presents spatial snapshots of this comparison for the years 1999, 2004, and 2013. The predicted fields capture the large-scale patterns and general distribution of the observed SST. While the overall agreement is encouraging, some regional differences are noticeable, particularly in the complex ocean currents near the northeast coast of the United States and Canada.

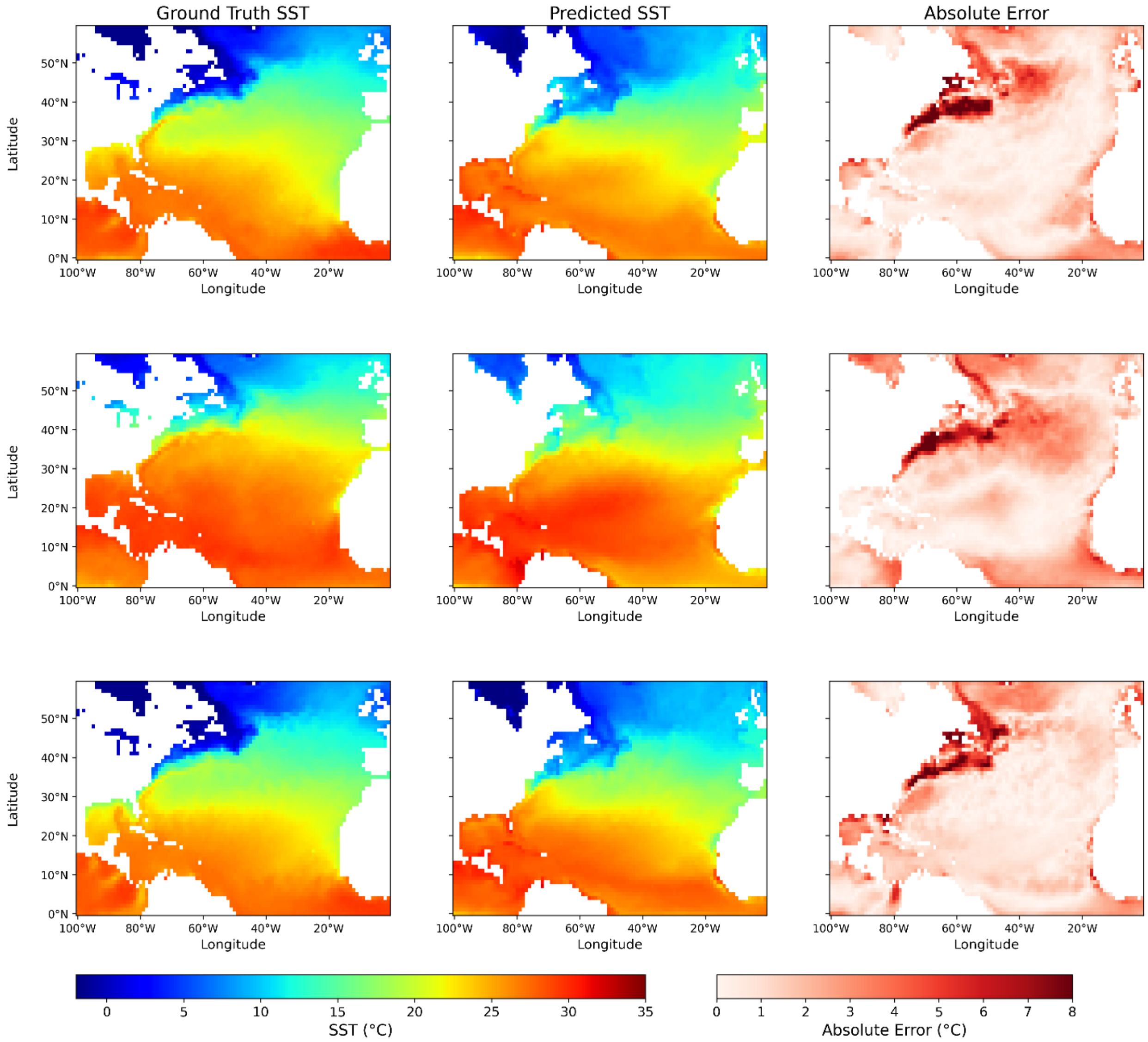

**Fig. 8.** Historical validation snapshots against ERA5 Reanalysis over the North Atlantic basin. Each column shows, from left to right, the reference ERA5 SST, predicted SST, and the absolute error between them. Each row corresponds to a different time within the validation period: 04/1999 (top), 10/2004 (middle), and 03/2013 (bottom).

A quantitative analysis of the prediction error (Predicted - Target) supports these visual findings. The multi-year basin-mean error was small across the key environmental variables, indicating little net basin-wide over- or under-predictions. For the representative case of SST, the mean error was −0.186 °C. Furthermore, the framework produced minimal bias when validated against ERA5 data for other parameters. For instance, it yielded a mean error of 0.317 °C ($\sigma$ = 2.282 °C) for air temperature, 0.001 hPa ($\sigma$ = 2.825 hPa) for sea-level pressure, and $4.198 \times 10^{-5} kg/kg$ ($\sigma$ $=8.817 \times 10^{-4} kg/kg$) for specific humidity. Comparable behavior holds for the remaining environmental drivers used by the hurricane model for which the corrected fields reproduce ERA5 large-scale patterns and exhibit near-zero basin-mean bias with stable dispersion. This consistent low-bias performance across multiple parameters supports the robustness of the correction step when applied to reanalysis data. To further assess the historical validation, the correction recovered by the framework was compared directly with the correction required to map GFDL-CM4 to ERA5 over the 1997–2014 hindcast period. The required correction was defined as the difference between ERA5 and GFDL-CM4, whereas the recovered correction was defined as the difference between the corrected field and GFDL-CM4. As shown in Fig. 9(a), the basin-mean monthly bias series exhibit strong temporal agreement, with a correlation coefficient of $r = 0.866$ and a bias-series RMSE of 0.367 °C. The correction also improves the overall agreement with ERA5, as summarized in Fig. 9(b). The RMSE decreases from 3.054 °C for the raw GFDL-CM4 field to 2.446 °C after correction, while the MAE decreases from 2.108 °C to 1.779 °C. The magnitude of the mean bias is reduced from 1.684 °C to 0.186 °C, corresponding to reductions of 19.9%, 15.6%, and 89.0% in RMSE, MAE, and mean-bias magnitude, respectively. These results indicate that the learned relationship captures the large-scale temporal evolution of the GCM bias while substantially reducing its systematic component. Regional discrepancies nevertheless remain, as illustrated by the spatial comparisons in Fig. 8.

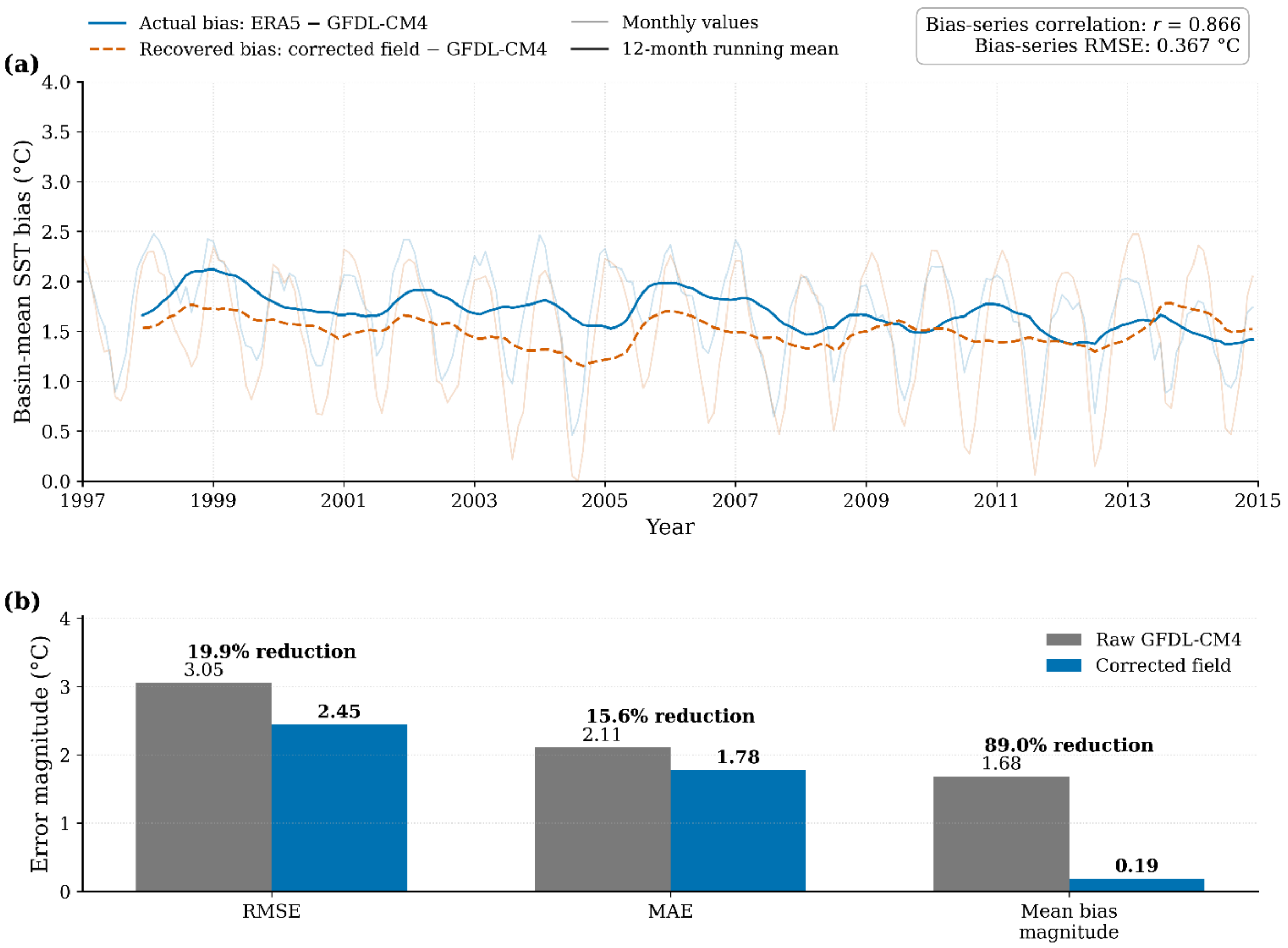

**Fig. 9.** Historical validation of the transferred GCM-derived mapping against ERA5 for 1997–2014. (a) Basin-mean temporal evolution of the actual correction required to map GFDL-CM4 to ERA5, $ERA5 - GFDL$, and the correction recovered by the proposed framework, $\widehat{ERA}5 - GFDL$. Thin curves represent monthly values and thicker curves represent 12-month running means; the reported correlation and RMSE are calculated from the original monthly series. (b) Comparison of the raw GFDL-CM4 and corrected fields against ERA5 in terms of RMSE, MAE, and mean-bias magnitude.

### 3.2 Simulation results of wind

Figure 10 shows the frequency distributions of simulated wind speeds for the historical climate (ERA5) and projected future climate scenarios across six study locations, generated using the proposed data-driven bias-correction framework. The near-future scenario corresponds to the simulation period 2024-2059, while the far-future scenario represents the simulation period 2060-2095. Overall, the results indicate a general trend towards higher wind speeds in the far-future scenario compared to near-future scenario, which is consistent with the anticipated intensification of hurricanes due to rising sea surface temperatures from 2024-2059 to 2060-2095. However, the magnitude of this increase varies notably among locations. Moreover, regions historically prone to intense hurricane activity, such as Miami, exhibit higher wind speed values compared to less vulnerable locations like Halifax. A more quantitative assessment of these changes will be conducted upon the determination of return periods.

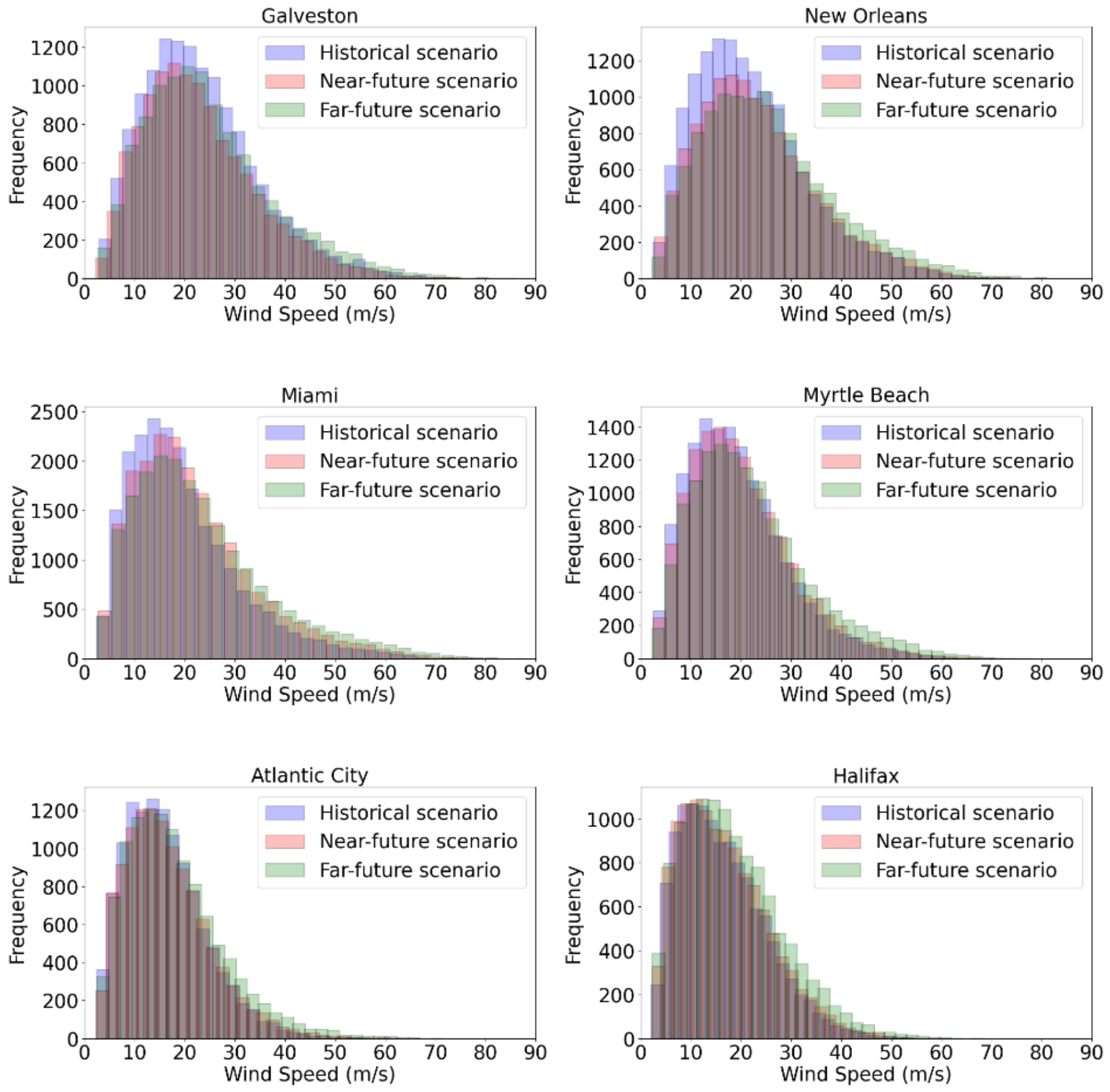

**Fig. 10.** Simulated wind speed for the historical and future climate scenarios at six locations.

A total of eleven return periods were generated for each of the selected locations. It should be noted that a track model similar to the one validated for historical scenarios by Emanuel(Emanuel, 2017) and Lin et al.(Lin et al., 2023) was employed in this study. As such, a revalidation of the model was deemed unnecessary. The findings of this study underscore the substantial impact of future climate scenarios on wind speeds, with significant variations observed across the six locations analyzed. These variations can be attributed to the disproportionate changes in hurricane intensity and frequency affecting coastal cities. To quantify the magnitude of these changes, the percentage change in wind speeds for MRIs was calculated by comparing future scenarios (2024-2059 and 2060-2095) to the historical period (1979-2014) across all locations. The results provide valuable insights into how wind climates may evolve over time, as shown in Fig. 11 and Table 1.

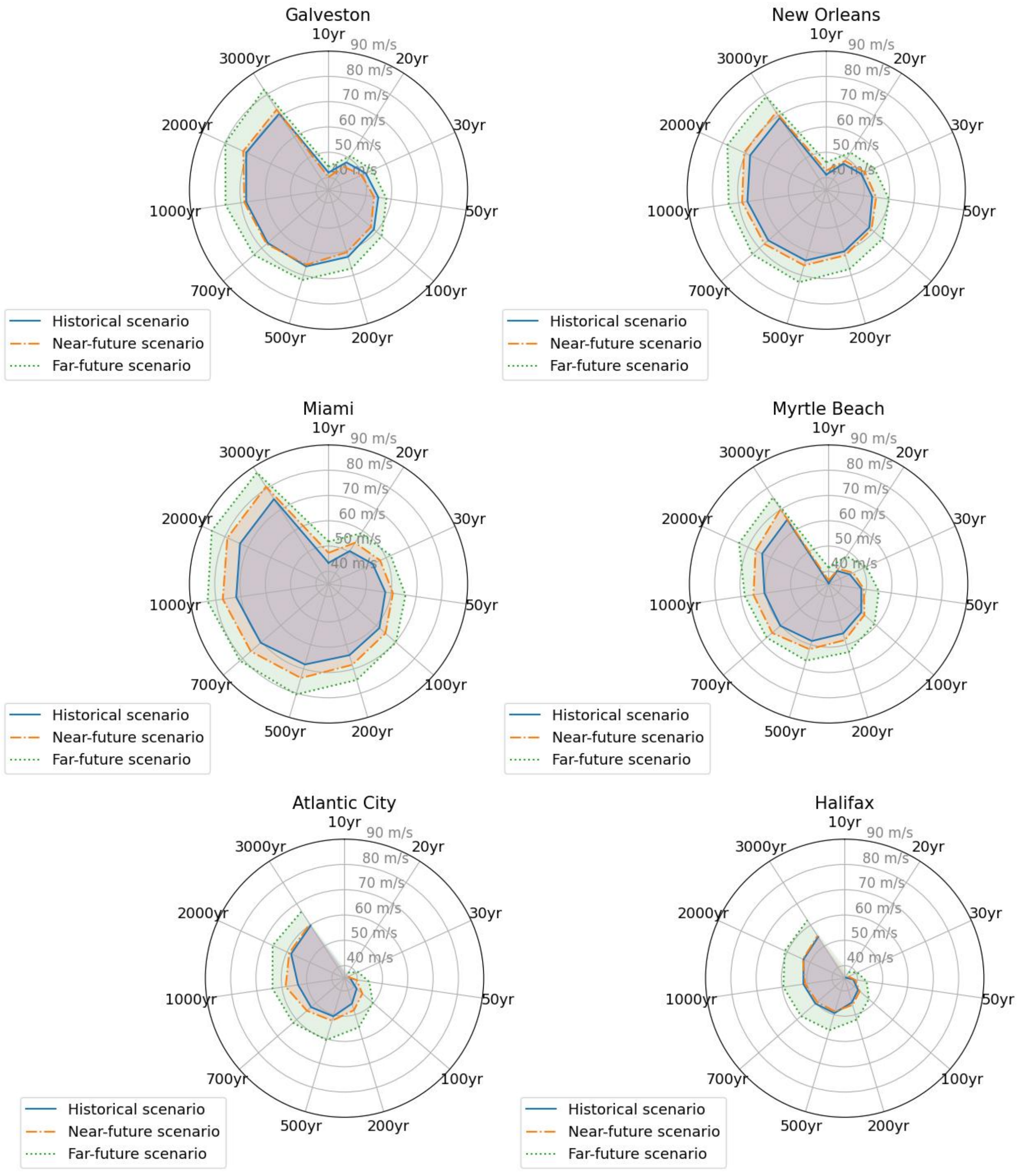

**Fig. 11.** Return period of wind speed for historical and future climate scenarios at six locations, simulated using the proposed bias-correction technique.

**Table 1** Percentage change in wind speeds for selected four return periods across locations compared to historical data (Scenario 1 = near-future scenario and Scenario 2 = far-future scenario)

| MRI / Location | 50 years | | 100 years | | 500 years | | 1000 years | |
|---|---|---|---|---|---|---|---|---|
| | Scenario 1 | Scenario 2 | Scenario 1 | Scenario 2 | Scenario 1 | Scenario 2 | Scenario 1 | Scenario 2 |
| Galveston | -3.13 | 5.89 | -2.55 | 7.00 | -0.95 | 8.51 | 1.11 | 12.17 |
| New Orleans | 2.93 | 12.55 | 1.99 | 11.96 | 3.17 | 14.02 | 3.12 | 11.20 |
| Miami | 5.23 | 13.95 | 5.11 | 14.04 | 8.19 | 18.09 | 7.46 | 15.93 |
| Myrtle Beach | 2.01 | 14.46 | 3.16 | 13.36 | 5.78 | 13.82 | 7.23 | 12.98 |
| Atlantic City | 6.36 | 19.73 | 7.23 | 20.44 | 3.57 | 19.29 | 9.30 | 19.07 |
| Halifax | 1.81 | 13.86 | 2.06 | 13.45 | -1.39 | 14.23 | -1.29 | 15.52 |

For Galveston, the near-future scenario predicts a slight decrease in wind speeds for the return periods (50, 100 and 500 years), with the most pronounced reduction observed at the 50-year return period (-3.13%). However, the far-future scenario forecasts a notable increase in wind speeds, with the most significant rise at the 1000-year return period (+12.17%), indicating a potential shift towards higher wind speeds as the century progresses. In New Orleans, wind speeds are generally projected to increase in both scenarios, with the far-future scenario showing more substantial rises across all return periods, particularly at the 50-year (+12.55%) and 500-year (+14.02%) return periods. Miami is projected to experience consistent increases in wind speeds across all return periods in both scenarios, with the far-future scenario having a more pronounced impact. The highest increases are observed for the 500-year (+18.09%) and 1000-year (+15.93%) return periods, exacerbating the risk of high-wind events in this hurricane-prone region. Myrtle Beach is also expected to see an overall increase in wind speeds, with the far-future scenario projecting the most significant rise at the 50-year return period (+14.46%), and similar increases observed for the longer return periods. In Atlantic City, wind speeds are projected to rise steadily across all return periods, with the far-future scenario showing particularly high increases (+19.73% for the 50-year and +20.44% for the 100-year return periods). Halifax is expected to experience moderate wind speed increases in the near-future scenario, but the far-future scenario forecasts a more substantial rise, especially at the 50-year return period (+13.86%). Interestingly, the 1000-year return period shows a slight decrease in the near-future scenario (-1.29%), yet the far-future scenario still reflects a rise (+15.52%), suggesting potential long-term intensification of wind speeds. Overall, the far-future scenario (2060-2095) tends to show more pronounced increases in wind speeds compared to the near-future scenario (2024-2059), particularly in regions like Atlantic City, Miami, and Myrtle Beach. This observation indicates that the latter half of the century may experience more severe wind hazards, which could have significant implications for wind-resistant building design, disaster preparedness, and climate resilience strategies across the studied regions (Wu et al., 2026).

The percentage differences between the data-driven bias-corrected framework and GFDL-CM4 data for the 2024-2059 period reveal distinct trends across various cities, as shown in Fig. 12. In Galveston, the differences are predominantly negative, with -16.37% for the 10-year return period, suggesting less intense storm events compared to GFDL-CM4. This trend persists but becomes less pronounced at longer return periods, with a smaller difference of -5.21% for the 1000-year period. New Orleans similarly shows negative differences, such as -8.25% for the 10-year and -4.04% for the 100-year return periods. Miami follows this trend, with differences of -8.23% for the 10-year and -4.69% for the 100-year return periods, although longer return periods (such as 700 years) show positive values, indicating more intense storms predicted by the data-driven bias-correction framework. Myrtle Beach continues to show negative differences for most return periods, including -2.89% for the 10 years and -2.84% for the 100 years. In contrast, Atlantic City and Halifax display positive differences across all return periods, with Atlantic City showing an increase of 6.55% for the 10 years and 9.09% for the 100 years, and Halifax seeing even larger positive differences, such as 9.63% for the 10 years and 9.28% for the 100 years. These positive differences suggest the data-driven framework predicts more intense storm events in these cities than GFDL-CM4 for the 2024-2059 period. For the 2060-2095 period, Galveston continues to show mostly negative differences, though less extreme, with -8.98% for the 10-year and -3.26% for the 100-year return periods. New Orleans shifts from -3.43% for the 10-year to a positive 8.14% for the 3000-year return period, indicating more intense storms predicted by the data-driven approach. Similarly, Miami shows a small negative difference of -0.99% for the 10-year but a positive 5.9% for the 1000-year return period. Myrtle Beach, Atlantic City, and Halifax exhibit large positive differences, with Myrtle Beach seeing a 6.63% increase for the 3000-year return period, Atlantic City showing 15.88% for the 50-year and 15.29% for the 100-year return periods, and Halifax with 18.7% for the 50 years and 18.26% for the 100 years. Overall, the bias-corrected simulations produce lower wind-speed return levels than simulations driven by the original GFDL-CM4 fields for most locations during 2024–2059, except Atlantic City and Halifax, where positive differences occur across the considered return periods. During 2060–2095, the bias-corrected simulations generally produce higher wind-speed return levels at most locations, with Galveston and the shorter return periods in New Orleans and Miami as exceptions. These location- and return-period-dependent differences demonstrate that bias correction of the environmental drivers can materially alter projected wind hazards.

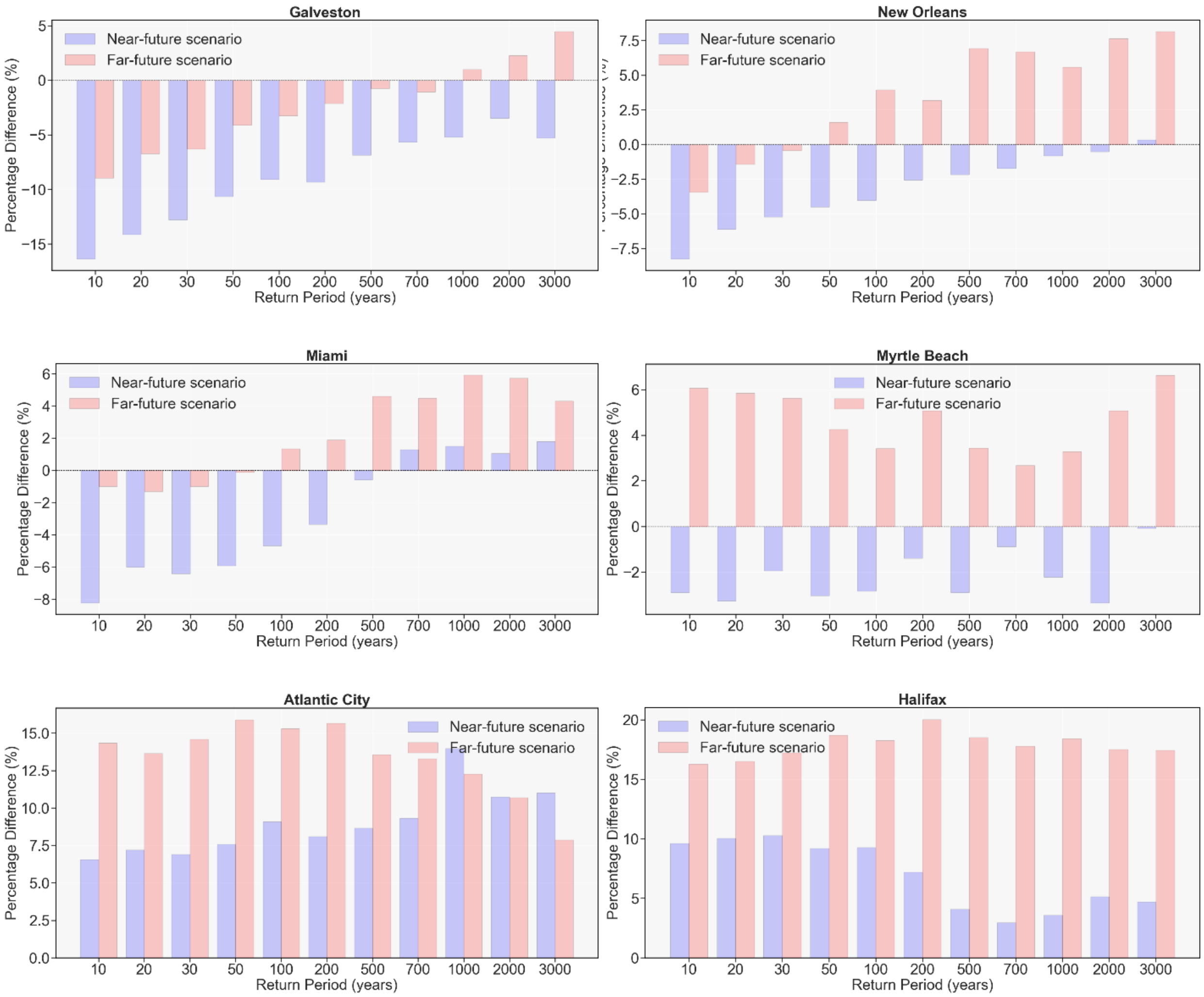

**Fig. 12.** Comparison of percentage differences between the data-driven bias-corrected framework and GFDL-CM4 data in wind speed projections for selected cities and return periods.

### 3.3 Simulation results of rain

The return periods of simulated rain rates under historical climate data (ERA5) and projected future climate scenarios were generated for the same six locations using the proposed data-driven bias-corrected framework. For consistency, only the percentage change in rain rates compared to historical data is presented in Table 2, along with the percentage change between the typical climate change assessment approach and the proposed framework in Fig. 13. The analysis reveals notable differences in rain rate projections across locations and return periods, reflecting varied impacts of climate change on extreme rainfall. Both the near-future scenario and the far-future scenario generally show increased rainfall intensities, with the far-future scenario consistently exhibiting a more substantial rise, indicating a stronger projected increase under the far-future climate state. Significant-change locations, such as Miami and Atlantic City, exhibit the largest percentage increases under the far-future scenario across all return periods, with Miami reaching a peak increase of 41.66% at the 1000-year return period, suggesting a potential for significantly intensified rainfall in future climate scenarios. Moderate-change locations like New Orleans show increases around 6.36% under the near-future scenario for the 100-year period, while lower-change locations like Galveston see only modest changes or even slight decreases for shorter return periods, as seen with a -3.20% change at the 50-year period under the near-future scenario. Across all locations, shorter return periods (50-100 years) show more conservative rain rate increases, often below 20% in the near-future scenario, with few exceptions, such as 13.51% increase for the 50-year period in Miami. However, longer return periods (500-1000 years) exhibit more pronounced changes, particularly in the far-future scenario, where Atlantic City reaches a 45.81% increase for the 500-year return period, suggesting intensified rainfall events in the long term. The near-future scenario shows more moderate rain-rate increases, generally below 20% for shorter return periods, whereas the far-future scenario exhibits substantially larger increases across locations and return periods, with 1000-year rain-rate increases exceeding 40% in Atlantic City and Miami. These findings highlight the importance of regional variability and return-period dependence in assessing the intensifying impacts of climate change on extreme rainfall, particularly in hurricane-prone coastal cities.

**Table 2** Percentage change in rain rates for selected four return periods across locations compared to historical data (Scenario 1 = near-future scenario and Scenario 2 = far-future scenario)

| MRI / Location | 50 years | | 100 years | | 500 years | | 1000 years | |
|---|---|---|---|---|---|---|---|---|
| | Scenario 1 | Scenario 2 | Scenario 1 | Scenario 2 | Scenario 1 | Scenario 2 | Scenario 1 | Scenario 2 |
| Galveston | -3.20 | 10.34 | -1.79 | 12.14 | -0.79 | 14.56 | 0.21 | 20.24 |
| New Orleans | 5.96 | 24.91 | 6.36 | 25.26 | 5.27 | 30.34 | 9.32 | 26.11 |
| Miami | 13.51 | 34.99 | 11.58 | 32.80 | 13.36 | 38.75 | 17.94 | 41.66 |
| Myrtle Beach | 2.22 | 22.74 | 1.74 | 23.60 | 0.97 | 22.75 | 2.35 | 23.99 |
| Atlantic City | 9.35 | 41.31 | 8.72 | 40.04 | 17.93 | 45.81 | 15.68 | 43.93 |
| Halifax | 6.97 | 37.18 | 9.20 | 37.05 | 9.22 | 37.72 | 9.43 | 37.90 |

The percentage differences between the data-driven bias-corrected framework and GFDL-CM4 data for the near-future and far-future scenarios periods reveal distinct trends across various cities, as shown in Fig. 13. For instance, Galveston is projected to experience a substantial decrease in rain rate, with a 27.6% reduction for a 10-year return period in 2024-2059 and a 17.6% decrease in 2060-2095. Conversely, Atlantic City is expected to see a substantial increase, with a 9.4% rise for a 10-year return period in the near-future period and a 22.1% increase in the far-future period. New Orleans exhibits a more nuanced trend, transitioning from a decrease in the near-future period to an increase in the far-future period especially for longer return periods. Miami shows in general a consistent decrease, with a 15% reduction for a 10-year return period in 2024-2059 and a 3.7% decrease in 2060-2095. Myrtle Beach experiences a mix of increases (far-future period) and decreases (near-future period), while Halifax consistently shows increases, particularly for longer return periods. These variations highlight the complex nature of climate change impacts and the need for localized assessments to inform adaptation and mitigation strategies.

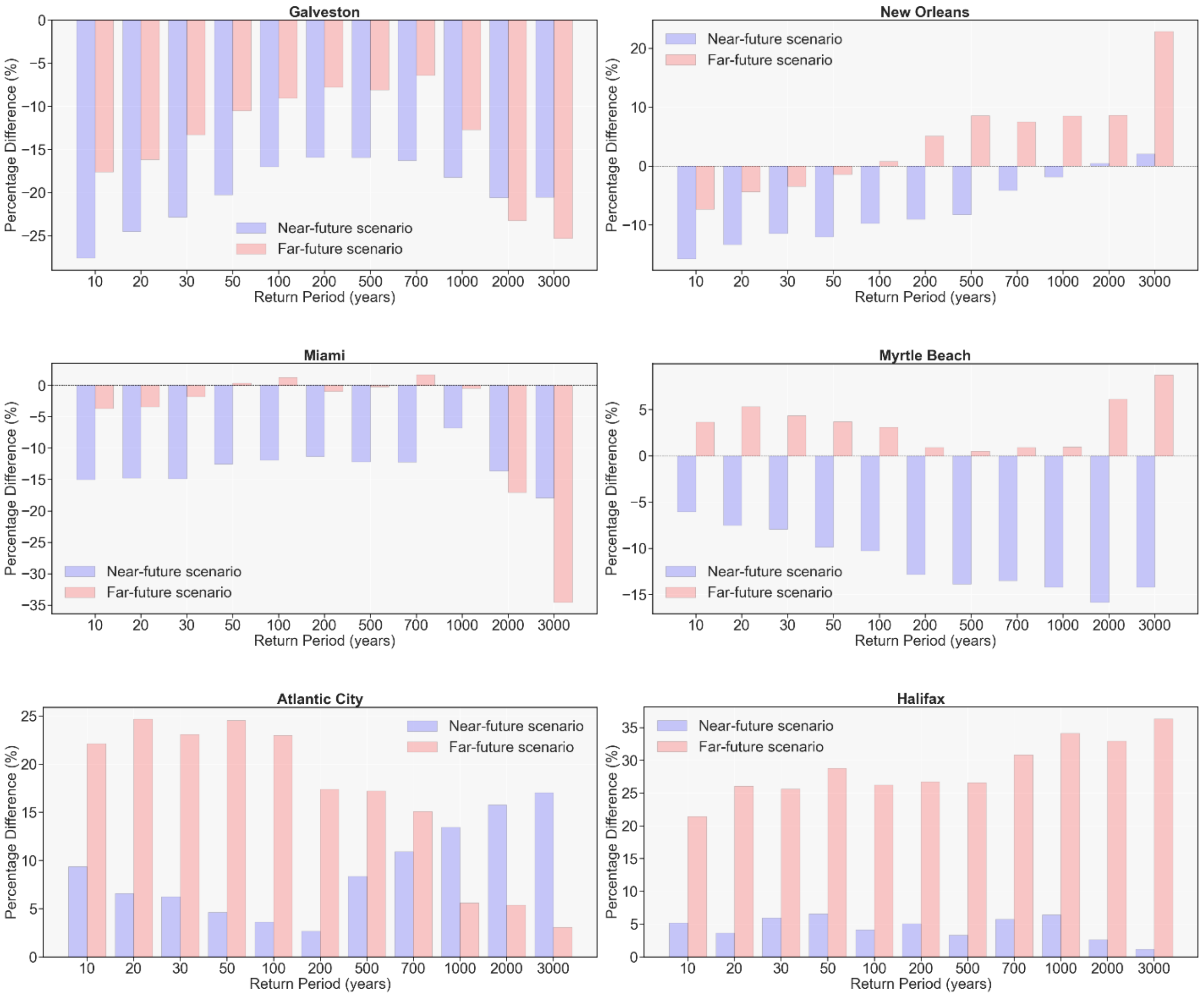

**Fig. 13.** Comparison of percentage differences between the data-driven bias-corrected framework and GFDL-CM4 data in rain rate projections for selected cities and return periods.

## 4. DISCUSSION

This study introduces a novel data-driven bias-correction framework designed to address the challenge of evolving biases in GCMs for future hurricane risk assessment. The primary focus of this work is the introduction and rigorous validation of this methodology. The framework is applied across the entire North Atlantic basin to generate the corrected environmental fields, and the subsequent analysis of wind and rain hazards at six selected cities is intended as a "proof-of-concept" to demonstrate the significant impact that this form of bias correction can have on local risk profiles. The high spatial dimensionality of GCM output data, encompassing variables such as monthly averaged sea surface temperature, mean sea level pressure, temperature and specific humidity at various pressure levels, and potential intensity – all required in the simulation of hurricane genesis, translation, and intensity – presents significant challenges for both computational efficiency and model interpretability. To address these challenges, dimensionality reduction is crucial. We apply dimensionality reduction to historical and projected data from the GFDL-CM4 model under the high-emissions SSP5-8.5 across three periods: 1979–2014, 2024–2059, and 2060–2095. This allows us to examine both short- and long-term climate effects. A surrogate model is then trained to map reduced GCM historical data to projected data, enabling the generation of future climate scenarios using ERA5 reanalysis data.

The application of our bias-correction framework to six coastal locations reveals a significant impact on projected hurricane activity compared to the raw GCM output. A quantitative analysis of storm occurrence frequency shows our framework projects a different number of storms over the 10,000-year simulation period, with the nature of this change varying by location. For instance, our approach leads to a notable reduction in the projected storm frequency for Miami and Atlantic City by approximately 11% and 12%, respectively, in the far-future scenario. Beyond storm frequency, the framework also influences the simulated storm trajectories. At the Miami site, for example, a distinct shift in storm paths is observed, with an increased number of storms moving toward the northwest and northeast compared to the historical scenario. This directional shift appears less pronounced in the far-future scenario, suggesting potential variations in climate change impacts over time. This behavior is relevant because GCMs are known to exhibit substantial regional biases in simulated tropical-cyclone frequency, which can manifest as either over- or under-prediction depending on the specific model and geographic basin (Strazzo et al., 2013; Roberts et al., 2020). These differences are consistent with the influence of regional GCM biases on simulated hurricane activity and demonstrate that bias correction of the environmental drivers can materially alter the resulting hazard estimates. While a comprehensive, basin-wide analysis using metrics like track density maps would be valuable, the primary focus of this study is to introduce and validate the bias-correction methodology. Presenting the quantitative impact on these specific case studies serves as a clear proof-of-concept to demonstrate the significance of the framework's effect.

These observed deviations result in substantial differences in wind speed and rain rate return periods between historical and future climates. Notably, wind speeds for 500-year and 1000-year return periods in Miami are projected to increase by +18.09% and +15.93%, respectively, while rain rates rise by +38.75% and +41.66%. These findings underscore the increasing risk of high-wind and heavy-rainfall events in hurricane-prone regions. While both the data-driven bias-corrected framework and the GFDL-CM4 model generate generally similar storm track patterns, some differences emerge, particularly in the near-future scenario, where the data-driven bias-corrected framework produces a higher number of northeast-directed storm tracks compared to the GFDL-CM4 model. These discrepancies highlight the influence of different modeling approaches in capturing storm behavior under changing climate conditions. Accordingly, comparisons with GFDL-CM4 data reveal spatial and temporal shifts in storm intensity predictions, with our framework forecasting more intense storms in most locations during 2060–2095. Exceptions are observed for lower return periods in cities such as Galveston and New Orleans. Our results emphasize the critical need for adaptive bias correction techniques and proactive planning to mitigate the growing risks posed by climate change-induced hurricane intensification.

The proposed mapping assumes that the relationship between historical and projected GCM climate states contains transferable information about the corresponding evolution of the real climate system. The historical hindcast provides evidence supporting this assumption over an independent period. Applying a relationship learned exclusively from GFDL-CM4 to the historical ERA5 state reduces the SST RMSE relative to the raw GCM by 19.9%, the MAE by 15.6%, and the magnitude of the mean bias by 89.0%. In addition, the recovered basin-mean bias evolution closely follows the actual ERA5–GFDL-CM4 bias, with a temporal correlation of $r = 0.866$.

Nevertheless, the hindcast RMSE of 2.446 °C remains higher than the internal GCM-to-GCM testing errors, and the spatial comparisons indicate remaining regional discrepancies. Direct observational validation of the near- and far-future periods is inherently unavailable; accordingly, the future projections remain conditional on the climate-change signal simulated by GFDL-CM4 under SSP5-8.5 and on the transferability of the learned relationship. Future work should therefore assess the framework across additional GCMs and climate scenarios, examine broader basin-scale diagnostics such as track-density distributions, and investigate more advanced nonlinear dimensionality-reduction approaches, such as convolutional autoencoders (Naeini and Snaiki, 2024; Naeini et al., 2025).

## 5. CONCLUSION

This study introduces a novel data-driven bias-correction framework for hurricane risk assessment under climate change. Dimensionality reduction is first applied to historical and projected data from the GFDL-CM4 model under SSP5-8.5, considering the historical (1979–2014), near-future (2024–2059), and far-future (2060–2095) periods. A reduced-order surrogate is then trained to learn the relationship between the historical and projected GCM states. The learned relationship is subsequently applied to a historical state derived from ERA5 reanalysis, and the projected environmental fields are reconstructed using the POD basis associated with the corresponding future GCM period. The proposed framework, trained on the full North Atlantic basin, was applied to six coastal locations within this region to simulate historical and future climate conditions. The simulation results demonstrate substantial differences in wind-speed and rain-rate return periods between historical and future scenarios, particularly for longer return periods. For instance, Miami is projected to experience consistent increases in wind speeds and rain rates across all return periods in both scenarios, with the far-future scenario having a more pronounced impact. The highest increases in wind speed are observed for the 500-year (+18.09%) and 1000-year (+15.93%) return periods, significantly increasing the risk of high-wind events in this hurricane-prone region.

Similarly, the highest increases in rain rate are observed for the 500-year (+38.75%) and 1000-year (+41.66%) return periods, exacerbating the risk of heavy rainfall and potential flooding. These findings highlight the potential for more severe hurricane-related wind and rainfall hazards under future climate conditions, emphasizing the need for proactive adaptation measures to mitigate associated risks. In addition, the comparison with simulations driven by the original GFDL-CM4 fields shows that the bias-corrected simulations generally produce lower wind-speed return levels during the 2024-2059 period. However, this pattern is reversed in certain locations, including Atlantic City and Halifax, where the bias-corrected simulations produce higher wind-speed return levels. For the 2060-2095 period, the bias-corrected simulations generally produce higher wind-speed return levels than simulations driven by the original GFDL-CM4 fields at most locations. Exceptions include Galveston and, at lower return periods, New Orleans and Miami. These results highlight the need for careful consideration of both model biases and the evolving nature of climate change when assessing future hurricane risks.

## Acknowledgements

This work was supported by the Natural Sciences and Engineering Research Council of Canada (NSERC) [grant number CRSNG RGPIN 2022-03492].